\documentclass[aps,pre,twocolumn,superscriptaddress,nofootinbib,amsmath,amssymb,floatfix]{revtex4-2}

\usepackage{graphicx}
\usepackage{dcolumn}
\usepackage{bm}
\usepackage{xcolor}
\usepackage[colorlinks=true,linkcolor=blue!50!black,citecolor=blue!50!black,urlcolor=blue!50!black]{hyperref}

\newcommand{\rr}{\mathbf{r}}
\newcommand{\kk}{\mathbf{k}}
\newcommand{\avg}[1]{\langle #1\rangle}
\newcommand{\cA}{\mathcal{A}}
\newcommand{\cB}{\mathcal{B}}

\begin{document}

\title{Pattern selection and biomass shifts in a vegetation model with nonlocal competition and Allee activation}
\author{F. Dimatz}
\affiliation{Universidad Nacional del Sur, Bah\'{\i}a Blanca, Buenos Aires, Argentina}
\author{M. N. Kuperman}
 \affiliation{FiEstIn, Gerencia de F\'{\i}sica, Centro At{\'o}mico Bariloche}
 \affiliation{Instituto Balseiro, Universidad Nacional de Cuyo}
 \affiliation{Consejo Nacional de Investigaciones Cient\'{\i}ficas y T\'ecnicas, 8400 S. C. de Bariloche, Argentina}

\date{\today}

\begin{abstract}
We study a single-species vegetation model in which a strong Allee effect acts locally and competition acts through a nonlocal kernel, and we describe the kernel with the one-parameter family of $q$-Gaussians, which goes from compact support to power-law tails. The local activation introduced by the Allee effect allows a finite-wavelength instability even when the Fourier transform of the kernel is positive, so Gaussian and heavy-tailed kernels can produce patterns. We derive in closed form the coefficients of the hexagonal amplitude equations and verify them against independent numerical steady states. The quadratic coefficient $Q$ results from the balance between the curvature of the local growth and the competitive inhibition at the pattern scale, and at onset its sign depends only on the fraction of the local activation that diffusion removes at the critical wavenumber. The same coefficient controls the morphology, since spots, labyrinths and gaps appear with pairwise competition only as $Q$ changes sign, and the amplitude equations predict the choice between hexagons and stripes in every simulated state in which the cubic truncation holds. It also controls the total biomass, which increases for spots and decreases for gaps.
\end{abstract}

\maketitle

%==========================================================
\section{Introduction}\label{sec:intro}
%==========================================================

Vegetation in arid and semiarid regions often organizes in regular patterns of spots, labyrinths, stripes, or gaps in a vegetated matrix \cite{klausmeier1999,vonhardenberg2001,rietkerk2008}. The usual interpretation of these structures is a finite-wavelength instability in which short-range facilitation acts as an activator and long-range competition for water acts as an inhibitor \cite{lefever1997,barbier2008}. A minimal description considers a single biomass field and represents the interaction through a kernel $G$ that weights the density in a neighborhood of each point. Fuentes, Kuperman and Kenkre showed that nonlocal competition in a Fisher-type equation produces spatial structures \cite{fuentes2003,fuentes2004}, and Pigolotti, L\'opez and Hern\'andez-Garc\'{\i}a showed that, in this class of models, patterns require the Fourier transform $\hat G(k)$ of the kernel to take negative values \cite{pigolotti2007}. The same condition governs the vegetation model of Mart\'{\i}nez-Garc\'{\i}a \textit{et al.} \cite{martinezgarcia2013,martinezgarcia2014}. Gaussian and exponential kernels have $\hat G>0$, and the same happens with kernels with power-law tails, so none of them can produce patterns in competition-only models. This is a strong restriction, if we consider that fat tails are common in dispersal kernels \cite{kot1996} and that there is no reason to expect a sharp cutoff in interactions mediated by water and roots.

Previous works showed that an Allee effect in the growth function relaxes this restriction \cite{clerc2005,clerc2010,escaff2015,vandervoort2026}. By making the per-capita growth increase with density, the Allee effect supplies a positive local feedback that can compete with the nonlocal inhibition, and van der Voort, Mart\'{\i}nez-Garc\'{\i}a and Doelman identified a general mechanism in which local growth outpaces the susceptibility to competition, which allows patterns with kernels of non-negative transform \cite{vandervoort2026}. Their analysis is linear and one-dimensional, it considers kernels with compact support, and it leaves the amplitude equations and the two-dimensional morphologies for future work. The linear analysis, however, says nothing about the symmetry of the pattern. For the two-dimensional nonlocal Fisher--Kolmogorov equation, numerical studies report almost exclusively hexagonal arrays of spots \cite{silvano2025}, and Pinto-Ramos, Surendran and Mart\'{\i}nez-Garc\'{\i}a showed, through amplitude equations, that for linear local growth pairwise competition restricts the stable patterns to spots, so that stripes and gaps require higher-order competition \cite{pintoramos2026}. Amplitude equations for related nonlocal models appear in Ref.~\cite{garlaschi2021}.

In the present work we consider a minimal vegetation model with local cooperative growth and nonlocal competition, and we follow the consequences of the local activation from the linear instability to the selected pattern and to the total biomass. It is not our intention to describe a particular ecosystem, and we do not claim that real competition kernels belong to the family that we use. Our more modest goal is to show that a single quantity, the quadratic coefficient $Q$ of the amplitude equations, organizes both the morphology of the patterns and the change of the total biomass. We use the $q$-Gaussian family because one parameter takes the kernel from compact support through the Gaussian to power-law tails, and because its Fourier transform has a closed form.

We organize the paper as follows. In Sec.~\ref{sec:model} we present the model and the kernel family, and in Sec.~\ref{sec:linear} we analyze the linear stability. In Sec.~\ref{sec:wnl} we derive the amplitude equations and the condition that selects the morphology, and in Sec.~\ref{sec:biomass} we relate the total biomass to the same coefficient. Section~\ref{sec:sim} compares these predictions with numerical simulations, and in Sec.~\ref{sec:disc} we discuss the results. The Appendices contain the derivations and the numerical details.

%==========================================================
\section{Model and kernels}\label{sec:model}
%==========================================================

We consider the biomass density $u(\rr,t)$ of a single species. Without space, a logistic growth with a cooperative (Allee) factor reads
\begin{equation}
\frac{1}{u}\frac{du}{dt}=b\left(1-\frac{u}{K}\right)\frac{u}{u+c}-d,
\label{eq:allee}
\end{equation}
with birth rate $b$, mortality $d$, carrying capacity $K$, and a density scale $c$ that sets the strength of the cooperative factor. The factor $u/(u+c)$ increases with density for $u\lesssim c$, which represents facilitation among individuals (shading, improved infiltration, pollination), and it makes the per-capita growth rate an increasing function of $u$ at low density. For $u\to0$ the per-capita rate tends to $-d<0$, so a population below a threshold density declines to extinction. Measuring $u$ in units of $K$ and time in units of $b^{-1}$ we obtain two dimensionless parameters, $\alpha=c/K$ and $\beta=d/b$.

To build a spatial model we make two assumptions. First, individuals disperse, and we represent this by diffusion with coefficient $D$ (in units of $\ell^2b$, with $\ell$ the interaction length defined below). Second, the two ingredients of the growth rate act at different ranges. Facilitation is a \emph{local} effect, whereas all the neighbors within some range compete for a shared resource such as water. We therefore keep the Allee factor local and replace the density in the competition factor $(1-u)$ by its nonlocal average $\bar u=G*u$,
\begin{equation}
\partial_t u=D\nabla^2u+F(u,\bar u),\quad
F(u,\bar u)=u\left[\frac{u(1-\bar u)}{u+\alpha}-\beta\right],
\label{eq:model}
\end{equation}
\begin{equation}
\bar u(\rr,t)=\int G(|\rr-\rr'|)\,u(\rr',t)\,d\rr' ,\qquad \int G\,d\rr=1 .
\end{equation}
Since $\int G=1$, homogeneous states have $\bar u=u$ and coincide with those of the local model, $u_0(1-u_0)=\beta(u_0+\alpha)$, i.e.
\begin{equation}
u_0^\pm=\tfrac12\Big[1-\beta\pm\sqrt{(1-\beta)^2-4\alpha\beta}\Big].
\label{eq:u0}
\end{equation}
The two roots are real for $\alpha<\alpha_s\equiv(1-\beta)^2/(4\beta)$, the saddle-node (fold) value at which they merge, and their distance $u_0^+-u_0^-=\sqrt{(1-\beta)^2-4\alpha\beta}$ vanishes at the fold. The extinction state $u=0$ is always stable, $u_0^-$ is the unstable Allee threshold, and $u_0^+\equiv u_0$ is stable to uniform perturbations. It is worth mentioning that $\alpha=c/K$ is the half-saturation scale of the cooperative factor and \emph{not} the Allee threshold, which is $u_0^-<(1-\beta)/2<1/2$ in units of $K$. Consequently $\alpha$ can be larger than one, the only restriction being $\alpha<\alpha_s$, and $\alpha_s>1$ whenever $\beta<3-2\sqrt2\simeq0.17$. We take $\beta=0.15$, for which $\alpha_s\simeq1.204$.

\subsection{The $q$-Gaussian kernel}\label{sec:qgauss}

We use the one-parameter family of radial $q$-Gaussian kernels \cite{tsallis1988},
\begin{equation}
G_q(r)\propto e_q(-\kappa r^2)=\big[1-(1-q)\kappa r^2\big]_+^{\frac{1}{1-q}},
\label{eq:qgauss}
\end{equation}
where $[x]_+=\max(x,0)$ and $\kappa>0$ sets the scale. In $d$ dimensions the family has compact support of radius $R=[(1-q)\kappa]^{-1/2}$ for $q<1$, with the top-hat as $q\to-\infty$ and the parabolic kernel at $q=0$. For $q=1$ it is the Gaussian, $G\propto e^{-\kappa r^2}$, and for $q>1$ it has power-law tails $G\propto r^{-2/(q-1)}$. The kernel is normalizable for $q<1+2/d$ and has finite variance only for $q<1+2/(d+2)$, which in $d=2$ means $q<2$ and $q<3/2$, respectively.

Kernels with different tails have different shapes, so the comparison between values of $q$ requires fixing a measure of the interaction scale. We consider three measures: (N1) the root-mean-square radius $\ell=\avg{r^2}^{1/2}$, finite only for $q<3/2$ in $d=2$; (N2) the median radius $r_{1/2}$, which exists for all normalizable kernels and which we set to $\sqrt{\ln2}\,\ell$; and (N3) the parameter $\kappa=1/\ell^2$. In $d=2$ the three coincide for the Gaussian. For $q>1$ we also use (N4) a fixed width of the kernel core, $a=1$ in the notation below, so that only the tail exponent changes. Unless stated otherwise we take $\ell=1$ as the unit of length and use N1.

With $n=1/(1-q)$, $s=1/(q-1)$, $\nu=n+d/2$ and $\mu=s-d/2$, the Fourier transform normalized to $\hat G(0)=1$ reads (Appendix~\ref{app:ft})
\begin{align}
\hat G_q(k)&=\Gamma(\nu+1)\Big(\frac{2}{kR}\Big)^{\nu}J_\nu(kR),\quad(q<1),
\label{eq:Gq_compact}\\
\hat G_q(k)&=\frac{2^{1-\mu}}{\Gamma(\mu)}(ka)^\mu K_\mu(ka),\quad(q>1),
\label{eq:Gq_tail}
\end{align}
where $R$ and $a$ are the length scales of the compact and power-law kernels, which the normalization fixes. For N1, $R=\ell\sqrt{(2n+d+2)/d}$ and $a=\ell\sqrt{2(\mu-1)/d}$; for N2 in $d=2$, $R=r_{1/2}/\sqrt{1-2^{-1/(n+1)}}$ and $a=r_{1/2}/\sqrt{2^{1/(s-1)}-1}$; for N3, $R=[(1-q)\kappa]^{-1/2}$ and $a=[(q-1)\kappa]^{-1/2}$.

\begin{figure}[t]
\includegraphics[width=\columnwidth]{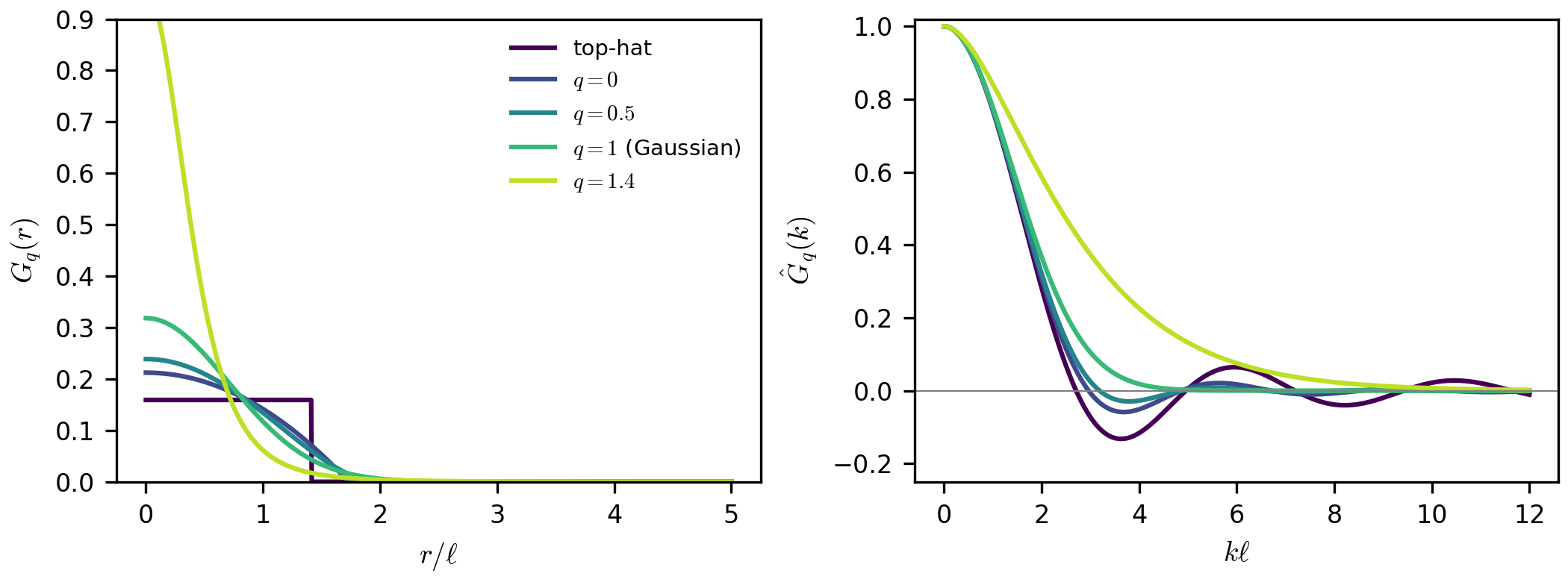}
\caption{$q$-Gaussian kernels in $d=2$ with the same rms radius $\ell=1$ (N1). Left: $G_q(r)$. Right: Fourier transform $\hat G_q(k)$. Compact kernels ($q<1$) have negative lobes whose depth decreases as $q\to1^-$, while the Gaussian ($q=1$) and the power-law kernel ($q=1.4$) have $\hat G>0$.}
\label{fig:kernels}
\end{figure}

The Bessel functions $J_\nu$ and $K_\mu$ in Eqs.~\eqref{eq:Gq_compact} and \eqref{eq:Gq_tail} appear because the Fourier transform of a radial function reduces to a one-dimensional (Hankel) integral \cite{watson1944}, and the function that appears decides the sign of $\hat G_q$, which is the property that matters for the instability. For $q<1$, $J_\nu$, the Bessel function of the first kind, oscillates for large argument, $J_\nu(x)\simeq\sqrt{2/\pi x}\cos(x-\nu\pi/2-\pi/4)$, so $\hat G_q$ has negative lobes and decays as $k^{-(\nu+1/2)}$. For the top-hat in $d=2$ we recover $2J_1(kR)/kR$, whose first minimum is $-0.13$, and the depth of the lobes decreases monotonically as $q\to1^-$ (Fig.~\ref{fig:kernels}). For $q>1$, $K_\mu$, the modified Bessel function of the second kind, is strictly positive and decays as $\sqrt{\pi/2x}\,e^{-x}$, so $\hat G_q>0$ for all $k$. For $q\to1$ both expressions tend to the Gaussian transform $e^{-k^2\ell^2/2d}$. It is also worth mentioning which part of the kernel controls which part of $\hat G$. The power-law tail $G\propto r^{-(2\mu+d)}$ produces a non-analytic term $\propto(ka)^{2\mu}$ at small $k$, while the decay $e^{-ka}$ at large $k$ reflects the width $a$ of the core. The tail then acts on long wavelengths and the core on short ones.

%==========================================================
\section{Linear stability analysis}\label{sec:linear}
%==========================================================

We consider a small perturbation of the homogeneous state, $u=u_0+\phi$. Since $F$ depends on $u$ and on $\bar u=G*u$, the linearization of Eq.~\eqref{eq:model} gives $\partial_t\phi=D\nabla^2\phi+F_u\phi+F_{\bar u}(G*\phi)$, with the derivatives evaluated at $(u_0,u_0)$. The Fourier modes $e^{i\kk\cdot\rr}$ decouple and grow or decay with rate (Appendix~\ref{app:lin})
\begin{equation}
\lambda(k)=-Dk^2+\cA+\cB\,\hat G(k),
\label{eq:lambda}
\end{equation}
where $\cA\equiv F_u=\alpha\beta/(u_0+\alpha)$ and $\cB\equiv F_{\bar u}=-u_0^2/(u_0+\alpha)$. Equation~\eqref{eq:lambda} contains two effects with opposite signs. On the one hand, $\cA>0$ is a \emph{local activation}, a direct consequence of the Allee effect: at the homogeneous state a small local excess of biomass increases the per-capita growth. On the other hand, $\cB\hat G(k)$ is a \emph{nonlocal inhibition} due to competition, and the term $-Dk^2$ damps short wavelengths. At $k=0$ the inhibition is maximal, $\hat G(0)=1$, and the uniform mode is stable below the fold,
\begin{equation}
\lambda(0)=\cA+\cB=-\frac{u_0\sqrt{(1-\beta)^2-4\alpha\beta}}{u_0+\alpha}<0 ,
\label{eq:lambda0}
\end{equation}
with a stability that weakens as the fold approaches, a fact that we use repeatedly below. A mode of wavenumber $k$ is unstable when $\lambda(k)>0$, i.e. when
\begin{equation}
D(u_0+\alpha)k^2+u_0^2\,\hat G(k)<\alpha\beta .
\label{eq:criterion}
\end{equation}
The onset occurs at a critical wavenumber $k_c$, where $\lambda(k_c)=\lambda'(k_c)=0$. We obtain the threshold $\alpha_c$ as the root of $\max_k\lambda(k;\alpha)=0$, and above onset we use $\sigma=\max_k\lambda$ as the distance to threshold in growth rate.

The mechanism follows directly from Eq.~\eqref{eq:lambda}. Without diffusion, since $\hat G(k)\to0$ at large $k$ for any integrable kernel, the activation $\cA$ would be unopposed at short scales and any $\alpha>0$ would give unstable modes of arbitrarily short wavelength. With diffusion, a kernel with $\hat G\ge0$ requires $\cA>Dk^2+|\cB|\hat G(k)$, so only wavenumbers $k<\sqrt{\cA/D}$ can grow, and $\lambda(0)<0$ excludes the uniform mode. A finite band of unstable wavenumbers can then exist without negative values of $\hat G$. The present model is a particular case of the mechanism of Ref.~\cite{vandervoort2026}, in which the dynamics reads $g(u)-s(u)\,G*c(u)$ and patterns with $\hat G\ge0$ require $\frac{d}{du}(g/s)>0$. With $g=u^2/(u+\alpha)-\beta u$, $s=u^2/(u+\alpha)$ and $c(u)=u$ we obtain $\frac{d}{du}(g/s)=\alpha\beta/u^2>0$ for any $\alpha>0$, and $s(u_0)\frac{d}{du}(g/s)|_{u_0}=\alpha\beta/(u_0+\alpha)=\cA$. The local activation of Eq.~\eqref{eq:lambda} is then exactly the excess of local growth over susceptibility of that reference, here produced by the Allee factor. In the competition-only model of Refs.~\cite{martinezgarcia2013,martinezgarcia2014} the local term is an inhibition and the instability requires $\hat G_{\min}<0$, with a critical competition strength $\delta_c\propto1/|\hat G|_{\min}$ that diverges as $q\to1^-$ (Appendix~\ref{app:lin}). In that model class no Gaussian or power-law kernel produces patterns, while in the present model they do.

\begin{figure*}[t]
\includegraphics[width=\textwidth]{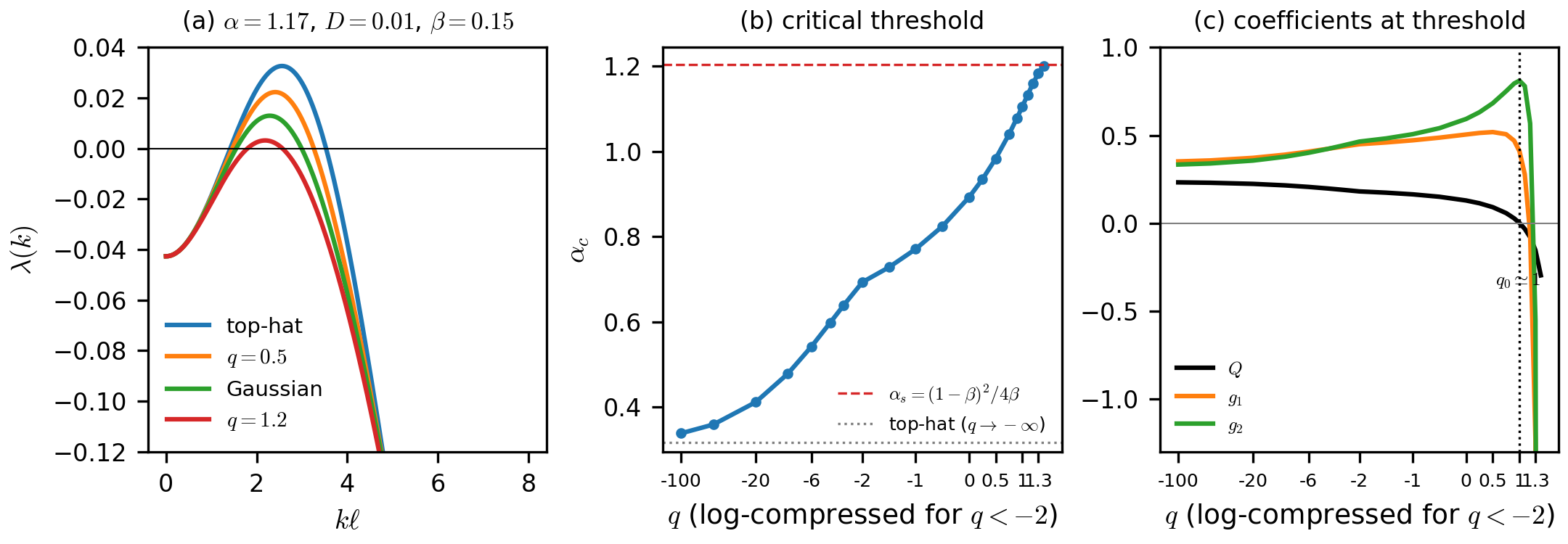}
\caption{Linear and weakly nonlinear analysis for $\beta=0.15$, $D=0.01$, $\ell=1$ ($d=2$), normalization N1. (a) Growth rate $\lambda(k)$ at $\alpha=1.17$ for the top-hat, $q=0.5$, the Gaussian and $q=1.2$. (b) Critical threshold $\alpha_c(q)$ (the horizontal axis is logarithmically compressed for $q<-2$). The dashed line is the existence limit $\alpha_s$ of the homogeneous state and the dotted line the top-hat value. (c) Quadratic ($Q$) and cubic ($g_1,g_2$) amplitude coefficients at threshold, Eqs.~\eqref{eq:Q} and \eqref{eq:g}. $Q$ changes sign at $q_0\simeq1.00$.}
\label{fig:linear}
\end{figure*}

\begin{figure*}[t]
\includegraphics[width=\textwidth]{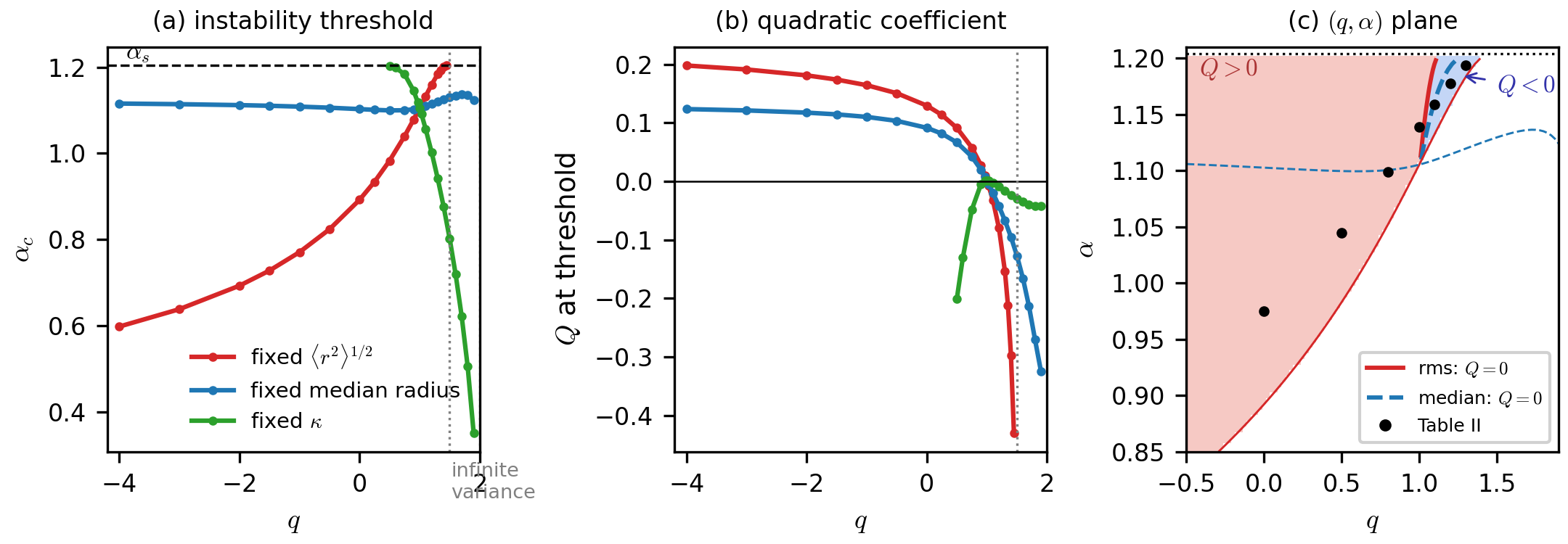}
\caption{Dependence on the measure of the interaction scale that we hold fixed. (a) Threshold $\alpha_c(q)$ for fixed rms radius (N1, defined for $q<3/2$), fixed median radius (N2) and fixed $\kappa$ (N3). The dashed line is $\alpha_s$. (b) Quadratic coefficient $Q$ at threshold. (c) Plane $(q,\alpha)$. The thin lines are the instability boundaries (solid, N1; dashed, N2), the shaded regions are the unstable region of N1 with $Q>0$ (red) and $Q<0$ (blue), and the thick lines are $Q=0$ at the most unstable wavenumber (solid, N1; dashed, N2). The dotted line is $\alpha_s$ and the dots mark the simulations of Table~\ref{tab:sim}.}
\label{fig:robust}
\end{figure*}

At $D=0.01$, $\beta=0.15$ and with the normalization N1, $\alpha_c$ increases monotonically with $q$, from $0.316$ for the top-hat to $1.105$ for the Gaussian, and $k_c$ decreases from $3.25$ to $2.46$ [Figs.~\ref{fig:linear}(a) and \ref{fig:linear}(b)]. The instability window closes near $q^*\simeq1.45$, where $\alpha_c$ reaches $\alpha_s$. This closure, however, is not a property of the tail. At fixed rms radius, $a=\ell\sqrt{\mu-1}\to0$ as $q\to3/2$ in $d=2$, because the variance constraint concentrates the kernel in a vanishing core. Competition then becomes effectively local at the scale of the pattern, $\hat G\simeq1$, and $\lambda(k)\to\lambda(0)<0$. When we fix another measure of the scale the effect disappears [Fig.~\ref{fig:robust}(a)]. With the median radius (N2) the instability persists up to $q=1.9$, including kernels of infinite variance, with $\alpha_c\simeq1.10$--$1.14$ and $k_c$ decreasing from $2.4$ at $q=1$ to $0.94$ at $q=1.9$, and with $\kappa$ fixed (N3) $\alpha_c$ even decreases with $q$. The trend of $\alpha_c$ with $q$ is then a property of the comparison protocol. What does not depend on the protocol is that no kernel of the family with $\hat G>0$ suppresses the instability, and that $k_c$ decreases as the tail becomes heavier.

%==========================================================
\section{Amplitude equations and morphology selection}\label{sec:wnl}
%==========================================================

The linear analysis gives the wavenumber $k_c$ at which the homogeneous state first becomes unstable, but not the symmetry of the pattern. Since the kernel and the equations are isotropic, every wavevector on the ring $|\kk|=k_c$ becomes unstable at the same parameter value, and the nonlinear terms decide among stripes, hexagons or other superpositions. Close to onset we can describe this competition with a few amplitudes \cite{cross1993,cross2009}. We summarize the steps here and give the full derivation for the present nonlocal model in Appendices~\ref{app:wnl} and \ref{app:amp}.

\subsection{Amplitude equations}\label{sec:ampeq}

Writing $u=u_0+\phi$ and expanding $F(u,\bar u)$ about $(u_0,u_0)$, the nonlinear terms up to third order are
\begin{align}
N_2&=a_2\phi^2+b_2\,\phi\,(G*\phi),\nonumber\\
N_3&=a_3\phi^3+b_3\,\phi^2(G*\phi),
\label{eq:N23}
\end{align}
where only products with at most one factor $G*\phi$ appear because $F$ is linear in $\bar u$. The Taylor coefficients $a_2=\tfrac12F_{uu}$, $b_2=F_{u\bar u}$, $a_3=\tfrac16F_{uuu}$ and $b_3=\tfrac12F_{uu\bar u}$ are
\begin{align}
a_2&=\frac{\alpha^2(1-u_0)}{(u_0+\alpha)^3},& b_2&=-\frac{u_0(u_0+2\alpha)}{(u_0+\alpha)^2},\nonumber\\
a_3&=-\frac{\alpha^2(1-u_0)}{(u_0+\alpha)^4},& b_3&=-\frac{\alpha^2}{(u_0+\alpha)^3}.
\label{eq:ab}
\end{align}
Since $a_2=\tfrac12\partial^2F/\partial u^2$ at fixed $\bar u$, it measures the curvature of the growth term as a function of the local density, with the competition held fixed. It is positive for any $\alpha>0$, because $u_0<1$, i.e. the Allee factor makes $F$ a convex function of $u$. In the same way, $b_2=\partial^2F/\partial u\,\partial\bar u<0$ measures how the competition modulates the response to the local density, and $a_3$, $b_3$ are the next order of the same expansion. Near onset we keep the three modes of a hexagonal lattice, $\kk_1+\kk_2+\kk_3=0$, $|\kk_j|=k_c$, with slowly varying complex amplitudes $A_j(t)$,
\begin{equation}
\phi=\sum_{j=1}^3\big(A_je^{i\kk_j\cdot\rr}+\text{c.c.}\big)+\phi_2 ,
\end{equation}
where $\phi_2$ contains the non-critical modes that the nonlinearity generates. The triad is special because the quadratic term couples two critical modes into the third one, $\kk_2+\kk_3=-\kk_1$, a resonance that stripes do not have. The non-critical modes (wavenumbers $0$, $\sqrt3k_c$ and $2k_c$) decay at rates of order one, follow the critical amplitudes adiabatically, and we eliminate them. Projecting on $e^{i\kk_1\cdot\rr}$ we obtain
\begin{equation}
\dot A_1=\sigma A_1+Q\bar A_2\bar A_3-g_1|A_1|^2A_1-g_2\big(|A_2|^2+|A_3|^2\big)A_1,
\label{eq:amp}
\end{equation}
and its cyclic permutations, with $\sigma=\lambda(k_c)$. The coefficient $Q$ is the strength of the quadratic resonance, $g_1$ the self-saturation of a mode and $g_2$ the cross-saturation between modes. The quadratic coefficient is
\begin{align}
Q&=2\big[a_2+b_2\hat G(k_c)\big]\nonumber\\
&=\frac{2\alpha^2(1-u_0)}{(u_0+\alpha)^3}-\frac{2\hat G(k_c)\,u_0(u_0+2\alpha)}{(u_0+\alpha)^2},
\label{eq:Q}
\end{align}
and the cubic ones are
\begin{align}
g_1&=-3P+Q\Big(\frac{2W_0}{\lambda_0}+\frac{W_2}{\lambda_2}\Big),\nonumber\\
g_2&=-6P+2Q\Big(\frac{W_0}{\lambda_0}+\frac{W_{\sqrt3}}{\lambda_{\sqrt3}}\Big),
\label{eq:g}
\end{align}
with $\lambda_0=\lambda(0)$, $\lambda_{\sqrt3}=\lambda(\sqrt3k_c)$ and $\lambda_2=\lambda(2k_c)$, all negative, $P=a_3+b_3\hat G(k_c)$, and $W_j=a_2+\tfrac12b_2[\hat G(jk_c)+\hat G(k_c)]$ for $j=0,\sqrt3,2$. The terms in $P$ are the direct cubic contributions, and the terms proportional to $Q$ come from the eliminated modes, with $W_j$ the coupling between a critical mode and a slaved mode of wavenumber $jk_c$ and $1/\lambda_j$ the inverse of its damping rate. The kernel enters only through $\hat G$ at $0$, $k_c$, $\sqrt3k_c$ and $2k_c$.

We must state two limits of validity. First, with $A_j=O(\varepsilon)$ and $\sigma=O(\varepsilon^2)$ the term $Q\bar A_2\bar A_3$ is $O(\varepsilon^2)$ and dominates the others unless $Q$ is itself small. As is customary, we retain $Q$ at its actual value. Second, the cubic truncation saturates the instability only if $g_1>0$ (and $g_1+2g_2>0$ for hexagons). Since $|\lambda(0)|$ becomes small close to the fold, the term $2QW_0/\lambda_0$ can grow and drive $g_1$ negative, and this happens at $q\simeq1.19$ with N1 and at $q\simeq1.35$ with N2 [Fig.~\ref{fig:linear}(c)]. Beyond these values the amplitude equations give the sign of $Q$, and with it the tendency towards spots or gaps, but they do not predict the saturated state.

\subsection{Independent verification of the coefficients}\label{sec:verif}

Equations~\eqref{eq:Q} and \eqref{eq:g} involve a long chain of algebra, so we checked them independently of the elimination of the slaved modes. We solved the stationary equation $D\nabla^2u+F(u,\bar u)=0$ with a Fourier--Galerkin Newton method on the periodic lattice of wavenumber $k_c$, keeping all harmonics up to $|\kk|\simeq3.5\,k_c$ (Appendix~\ref{app:verif}), and compared the amplitude of the first harmonic with Eq.~\eqref{eq:amp}, which gives $\sigma=g_1A^2$ for stripes and $\sigma+QA-(g_1+2g_2)A^2=0$ for hexagons. The stable hexagonal branch has a large amplitude, $A\simeq Q/(g_1+2g_2)$, outside the regime of the expansion, so we used the unstable small-amplitude branch, $A\simeq-\sigma/Q$, where the expansion holds. When we extrapolate to $\sigma\to0$, the numerical coefficients agree with the formulas to better than $1\%$ (Table~\ref{tab:verif}). This validates $Q$, $g_1$ and, through $g_1+2g_2$, also $g_2$.

\begin{table}[t]
\caption{Independent check of the cubic coefficients: formula, Eq.~\eqref{eq:g}, against Fourier--Galerkin steady states extrapolated to $\sigma\to0$ (N1, $\beta=0.15$, $D=0.01$). Stripes test $g_1$ and the small-amplitude hexagon branch tests $g_1+2g_2$. For $q=1$, where $Q\simeq0$, we use the stable branch and the accuracy is $\sim1\%$; for $q=1.2$ the extrapolation uses three points. Dashes: $g_1<0$.}
\label{tab:verif}
\begin{ruledtabular}
\begin{tabular}{rrrrr}
 & \multicolumn{2}{c}{$g_1$} & \multicolumn{2}{c}{$g_1+2g_2$} \\
$q$ & formula & numerical & formula & numerical \\
\hline
$-\infty$ & 0.345 & 0.345 & 1.000 & 0.999 \\
$0.5$ & 0.519 & 0.518 & 1.884 & 1.883 \\
$0.8$ & 0.498 & 0.498 & 2.031 & 2.026 \\
$1$ & 0.410 & 0.409 & 2.03 & 2.02 \\
$1.1$ & 0.278 & 0.275 & 1.839 & 1.827 \\
$1.2$ & -- & -- & 1.077 & 1.07 \\
\end{tabular}
\end{ruledtabular}
\end{table}

\subsection{What selects the morphology}\label{sec:morphtheory}

Hexagons ($A_j=A$ real) obey $\sigma A+QA^2-(g_1+2g_2)A^3=0$ and exist for $\sigma>-Q^2/[4(g_1+2g_2)]$, i.e. \emph{below} the linear threshold when $Q\ne0$. The bifurcation is subcritical, with bistability between the homogeneous and the hexagonal states. On the large-amplitude branch $A$ has the sign of $Q$, which gives maxima of $u$ on the lattice (spots) for $Q>0$ and minima (gaps) for $Q<0$. Stripes ($A_1\ne0$, $A_{2,3}=0$) have $|A_1|^2=\sigma/g_1$ and are stable only if $g_2>g_1$ and $\sigma>Q^2g_1/(g_2-g_1)^2$, while hexagons are stable for $\sigma<Q^2(2g_1+g_2)/(g_2-g_1)^2$ (Appendix~\ref{app:amp}). As $Q\to0$ the quadratic resonance disappears, hexagons lose their advantage, and the system favors stripes, which appear as labyrinths in an extended domain.

The difference with competition-only models lies in $Q$, which is the sum of two contributions with opposite signs. On the one hand, the curvature of the local growth contributes the positive term $2a_2$, independent of the kernel. A convex growth means that a region with a local excess of biomass grows more than a region with the same deficit decays, which favors peaks of biomass, i.e. spots. On the other hand, the term $2b_2\hat G(k_c)$ is negative when $\hat G(k_c)>0$. A peak of biomass also raises the competition felt in its neighborhood at the scale of the pattern, and this penalizes peaks and favors depressions, i.e. gaps. In the nonlocal Fisher--Kolmogorov equation the first effect is absent, $a_2=0$, so that $Q=2|b_2||\hat G(k_c)|$ is positive because the instability requires $\hat G(k_c)<0$, and gaps cannot appear \cite{pintoramos2026}. Pinto-Ramos \textit{et al.} recover gaps by adding three-body competition, which contributes to the quadratic coefficient with the sign that favors gaps. In the present model the pairwise term itself can favor gaps, because the local activation allows the instability with $\hat G(k_c)>0$, while the curvature of the growth opposes it. $Q$ then vanishes when
\begin{equation}
\hat G(k_c)=\frac{a_2}{|b_2|}=\frac{\alpha^2(1-u_0)}{u_0(u_0+\alpha)(u_0+2\alpha)}\equiv G^*(\alpha)>0 ,
\label{eq:Gstar}
\end{equation}
so that $Q=2|b_2|[G^*-\hat G(k_c)]$. The sign of $Q$ then depends on the \emph{value} of the kernel transform at the pattern wavenumber relative to a constant that depends only on the growth function. We obtain $Q>0$ (spots) if $\hat G(k_c)<G^*$ and $Q<0$ (gaps) if $\hat G(k_c)>G^*$. Van der Voort \textit{et al.} observed in one dimension that the sign of the kernel transform at the pattern wavenumber decides whether the competitive pressure concentrates in the gaps or within the vegetated patches \cite{vandervoort2026}. Equation~\eqref{eq:Gstar} turns this observation into a quantitative criterion for the morphology, in which the relevant comparison is not with zero but with the positive value $G^*$.

At the threshold this condition takes a particularly simple form. The marginal condition $\lambda(k_c)=0$ fixes $\hat G(k_c)=(\cA-Dk_c^2)/|\cB|$, and using $\beta=u_0(1-u_0)/(u_0+\alpha)$ we find $|\cB|G^*=\cA\,\alpha/(u_0+2\alpha)$. The condition $Q>0$ is then equivalent to
\begin{equation}
\frac{Dk_c^2}{\cA}>\frac{u_0+\alpha}{u_0+2\alpha} .
\label{eq:onset}
\end{equation}
The left-hand side is the fraction of the local activation that diffusion removes at the pattern wavenumber, and the right-hand side depends only on the growth function and lies between $1/2$ and $1$. At onset the kernel enters the morphology only through $k_c$. When diffusion removes most of the activation at the pattern scale, the instability needs a small or negative $\hat G(k_c)$ and the pattern is made of spots. When diffusion removes only a small part, the instability survives with a large $\hat G(k_c)$ and the pattern is made of gaps. For the top-hat $Dk_c^2>\cA$ at all the parameters that we examined, so only spots are possible. For the Gaussian at $\beta=0.15$, $D=0.01$ the two sides of Eq.~\eqref{eq:onset} are $0.601$ and $0.599$, which is why $Q\simeq0$ there [Fig.~\ref{fig:linear}(c)], and heavier tails, which lower $k_c$, give $Q<0$.

The value $q_0$ where $Q$ changes sign depends on the parameters and on the normalization. With N1 and $\beta=0.15$ we obtain $q_0\simeq1.00$ at threshold, and at fixed $\alpha=1.15$, $1.17$, $1.19$ we obtain $q_0\simeq1.04$, $1.06$, $1.09$ (N1) and $1.07$, $1.12$, $1.19$ (N2) [Fig.~\ref{fig:robust}(c)]. For $\alpha\lesssim1.1$ there is no sign change in the unstable range of N1, and $Q>0$ for every kernel that destabilizes the homogeneous state. The sequence spots $\to$ labyrinths $\to$ gaps driven by $q$ therefore requires parameters close to the fold, $\alpha_c/\alpha_s\gtrsim0.92$ for $\beta=0.15$.

%==========================================================
\section{Total biomass}\label{sec:biomass}
%==========================================================

The same coefficient that selects the morphology also controls how the pattern changes the total biomass. When we integrate Eq.~\eqref{eq:model} over a periodic domain, diffusion drops out and the total biomass $M=\int u\,d\rr$ obeys $dM/dt=\int F(u,\bar u)\,d\rr$. A stationary state therefore has $\avg{F}=0$, where $\avg{\cdot}$ is the spatial average. The mean growth must vanish even though the local growth does not, and since $F$ is not linear in $u$, the mean of $F$ differs from $F$ evaluated at the mean, and the mean density must shift away from $u_0$ to restore the balance. To second order in the amplitude the mean of $\phi$ is the slaved $k=0$ mode, and using $\avg{\delta u^2}=2\sum_j|A_j|^2$ and $\lambda(0)<0$ we obtain (Appendix~\ref{app:bio})
\begin{equation}
\avg{u}-u_0\simeq\frac{Q}{2|\lambda(0)|}\,\avg{\delta u^2}.
\label{eq:biomass}
\end{equation}
The pattern increases the total biomass for $Q>0$ (spots) and decreases it for $Q<0$ (gaps), and at $Q=0$ labyrinths are biomass-neutral at this order. Equation~\eqref{eq:biomass} is a Jensen-type consequence of $\avg F=0$ and of the curvature of $F$, and at this order the shift has the sign of the skewness, since for a hexagonal pattern the skewness is $\mathrm{sign}(A)\,2/\sqrt6\simeq\pm0.82$ and a stripe pattern has zero skewness. What Eq.~\eqref{eq:biomass} adds is the \emph{amplitude} of the shift. Together with the hexagonal branch of Eq.~\eqref{eq:amp}, $|A|=[|Q|+\sqrt{Q^2+4g\sigma}]/(2g)$ with $g=g_1+2g_2>0$ and $\avg{\delta u^2}=6A^2$, it predicts the shift from the parameters $(\alpha,\beta,D,q)$ alone. Since $|\lambda(0)|$ vanishes at the fold, we expect this prediction to deteriorate as the fold approaches.

%==========================================================
\section{Simulations}\label{sec:sim}
%==========================================================

\subsection{Setup and observables}\label{sec:methods}

We integrate Eq.~\eqref{eq:model} in $d=2$ on a periodic $L\times L$ domain with $N^2$ points, using a first-order semi-implicit pseudospectral scheme with implicit diffusion and explicit reaction,
\begin{equation}
\hat u^{n+1}=\frac{\hat u^n+\Delta t\,\hat F^n}{1+\Delta t\,Dk^2},
\label{eq:scheme}
\end{equation}
where we evaluate the nonlocal average by FFT with the analytical $\hat G(k)$. The runs use $\beta=0.15$, $D=0.01$, $\ell=1$, $L=32$, $N=192$, $\Delta t=0.25$ and $T=4000$, start from $u_0[1+10^{-3}\eta]$ with $\eta$ Gaussian white noise, and we ran six realizations per kernel. To place all kernels at the same distance from threshold in growth rate, we choose $\alpha$ for each kernel such that $\sigma=\max_k\lambda=6\times10^{-3}$. Appendix~\ref{app:num} contains the numerical details, and the Supplemental Material \cite{SM} the convergence tests.

We characterize each state by the biomass $m=\avg u/u_0$, the skewness of $u$, the Euler characteristic density $\chi/L^2$ of the set $\{u>\avg u\}$, the structure factor $S(\kk)=L^{-2}|\hat u(\kk)|^2$ with its dominant wavenumber $k^*$, and the hexatic order $\overline{|\psi_6|}$ of the maxima or minima. The skewness and $\chi$ are positive for spots, close to zero for labyrinths and negative for gaps. Since the threshold $\avg u$ that defines $\chi$ is a convention, we use $\chi$ only together with the skewness.

\subsection{The sequence of morphologies}\label{sec:morph}

\begin{figure*}[t]
\includegraphics[width=\textwidth]{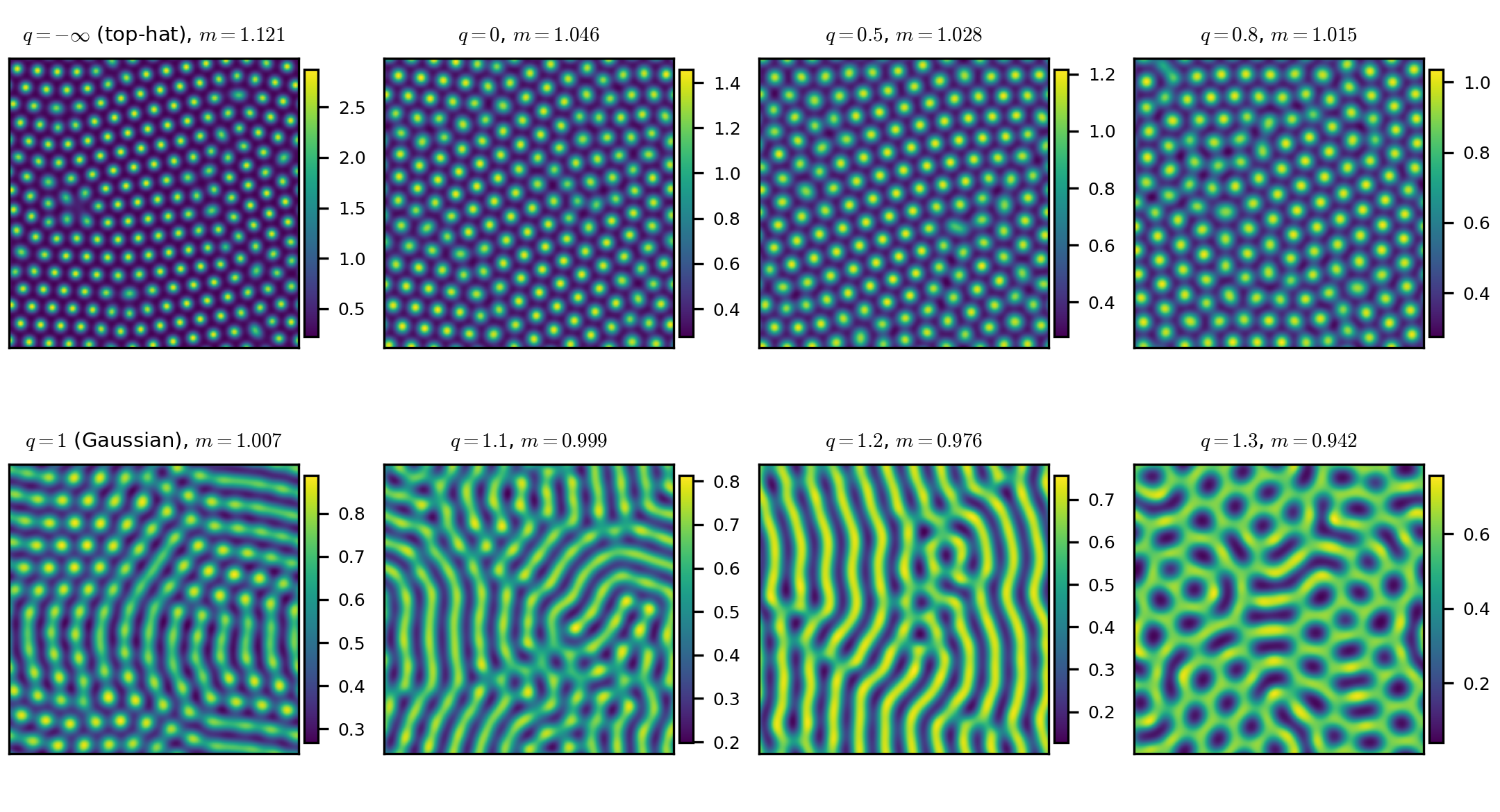}
\caption{Stationary biomass fields $u(\rr)$ at $T=4000$ for the kernels of Table~\ref{tab:sim} ($d=2$, $L=32$, realization 1), each panel with its own color scale and labeled with $q$ and the relative biomass $m$. With pairwise competition only we obtain spots ($q\le0.8$), a mixture of spots and stripes ($q=1$), a labyrinth ($q=1.1$) and gaps ($q\ge1.2$).}
\label{fig:snapshots}
\end{figure*}

\begin{figure}[t]
\includegraphics[width=\columnwidth]{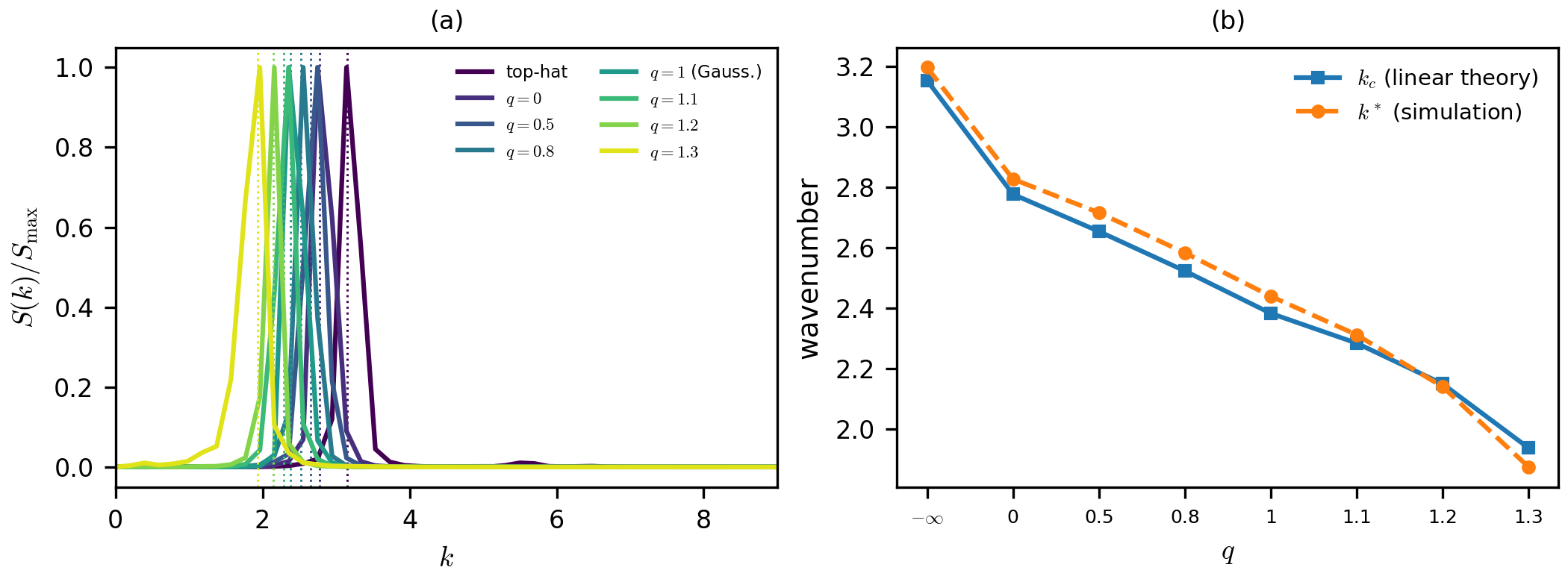}
\caption{(a) Angular-averaged structure factor for each $q$, normalized by its maximum. The dotted lines mark the linear prediction $k_c$. (b) Dominant wavenumber $k^*$ in the simulation compared with $k_c$ (realization 1).}
\label{fig:sk}
\end{figure}

The quadratic coefficient at threshold decreases monotonically with $q$ and changes sign at $q_0\simeq1.00$ [Fig.~\ref{fig:linear}(c)]. As we can see in Fig.~\ref{fig:snapshots} and Table~\ref{tab:sim}, the simulations follow the tendency of $Q$. For $q\le0.8$ ($Q>0$) the state is an array of spots with local hexagonal order ($\overline{|\psi_6|}\simeq0.8$--$0.9$, against $0.36$ for random points). At $q=1$ we observe a mixture of spots and stripes, with a large dispersion between realizations (skewness $0.29\pm0.09$), and at $q=1.1$ ($Q\simeq0$) a labyrinth with defects. For $q=1.2$ and $1.3$ ($Q<0$) the skewness and the Euler density are negative, and the pattern consists of low-biomass gaps in a vegetated matrix. For all the patterns the structure factor is a ring whose radius $k^*$ follows $k_c$ within the resolution $\Delta k\simeq0.2$ (Fig.~\ref{fig:sk}).

The amplitude equations also predict which state is stable where the cubic truncation holds. Evaluating $Q$, $g_1$ and $g_2$ at the simulated $\alpha$ and $\sigma=6\times10^{-3}$, the conditions of Sec.~\ref{sec:morphtheory} predict stable hexagons and unstable stripes for $q\le0.8$, and stable stripes and unstable hexagons for $q=1$ ($Q=0.011$) and $q=1.1$ ($Q=-0.023$). This agrees with the spots that we observe for $q\le0.8$, with the labyrinth at $q=1.1$, and with the stripe domains that coexist with spots at $q=1$. For $q\ge1.2$, $g_1<0$ and the truncation makes no prediction.

It is worth mentioning that the boundary between labyrinths and gaps is a crossover and not a sharp transition. For $q=1.2$ the skewness depends on the domain size and on the integration time in opposite directions: it becomes more negative in $L=64$ and less negative when we continue the runs to $T=16000$ \cite{SM}. Spot patterns, on the contrary, change by less than $3\%$ with $\Delta t$, $N$, $L$ and $T$.

\begin{table*}[t]
\caption{Summary of the simulations (N1, $\beta=0.15$, $D=0.01$), with $\alpha$ chosen such that $\max_k\lambda=6\times10^{-3}$. Means over six realizations at $t=4000$, with the sample standard deviation in parentheses in units of the last digit. $k_c$ and $Q$ are linear predictions at the simulated $\alpha$, and $k^*$ is the centroid of the structure-factor peak ($\Delta k=0.20$). $\overline{|\psi_6|}$ is the hexatic parameter of the maxima (spots) or minima (gaps); we do not report it for labyrinths.}
\label{tab:sim}
\begin{ruledtabular}
\begin{tabular}{rrrrrrrrrrr}
$q$ & $\alpha$ & $\alpha/\alpha_c$ & $k_c$ & $k^*$ & $Q$ & $m$ & skewness & $\chi/L^2$ & $\overline{|\psi_6|}_{\rm max}$ & $\overline{|\psi_6|}_{\rm min}$ \\
\hline
$-\infty$ & 0.538 & 1.70 & 3.15 & 3.18 & $+0.234$ & 1.124(3) & 1.32(1) & 0.226(1) & 0.90(4) & -- \\
$0$ & 0.975 & 1.09 & 2.78 & 2.82 & $+0.137$ & 1.045(2) & 1.01(1) & 0.178(4) & 0.88(5) & -- \\
$0.5$ & 1.045 & 1.06 & 2.65 & 2.72 & $+0.100$ & 1.029(1) & 0.92(1) & 0.165(1) & 0.84(3) & -- \\
$0.8$ & 1.099 & 1.04 & 2.52 & 2.58 & $+0.058$ & 1.015(1) & 0.78(2) & 0.144(2) & 0.79(4) & -- \\
$1$ & 1.139 & 1.03 & 2.38 & 2.43 & $+0.011$ & 1.006(1) & 0.29(9) & 0.038(17) & 0.67(6) & -- \\
$1.1$ & 1.159 & 1.02 & 2.28 & 2.31 & $-0.023$ & 0.999(1) & $-$0.02(2) & $-$0.002(6) & -- & -- \\
$1.2$ & 1.178 & 1.02 & 2.15 & 2.12 & $-0.070$ & 0.981(2) & $-$0.28(8) & $-$0.033(11) & -- & 0.64(6) \\
$1.3$ & 1.194 & 1.01 & 1.94 & 1.88 & $-0.144$ & 0.939(5) & $-$0.50(6) & $-$0.057(7) & -- & 0.63(6) \\
\end{tabular}
\end{ruledtabular}
\end{table*}

\subsection{What selects the morphology in the simulations}\label{sec:collapse}

\begin{figure*}[t]
\includegraphics[width=0.85\textwidth]{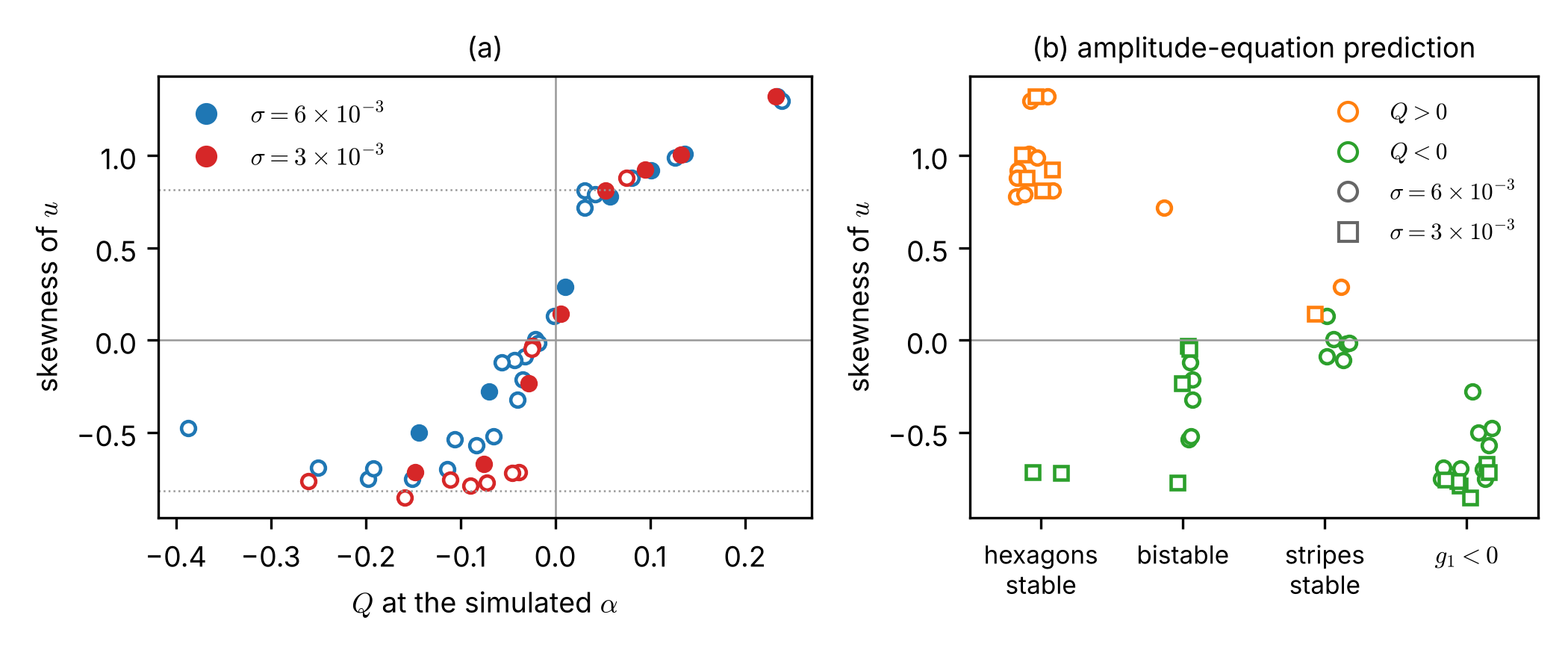}
\caption{Morphology of the $49$ simulated states, those of Table~\ref{tab:sim} and those of the Supplemental Material \cite{SM}. (a) Skewness of the stationary field against the quadratic coefficient $Q$ evaluated at the simulated $\alpha$, for $\sigma=\max_k\lambda=6\times10^{-3}$ (blue) and $3\times10^{-3}$ (red). Filled symbols: N1 with $\beta=0.15$; open symbols: N1 with $\beta=0.3$, N2, N3, N4 and N1 at other values of $D$. The dotted lines are the skewness $\pm2/\sqrt6$ of an ideal hexagonal pattern. (b) The same states grouped according to the stable states predicted by Eq.~\eqref{eq:amp} at the simulated $\alpha$ and $\sigma$: only hexagons, hexagons and stripes (bistable), only stripes, and $g_1<0$, where the cubic truncation does not apply. Circles: $\sigma=6\times10^{-3}$; squares: $\sigma=3\times10^{-3}$.}
\label{fig:collapse}
\end{figure*}

Along the sequence of Table~\ref{tab:sim}, $Q$ and $q$ change together, so this sequence alone cannot tell whether the morphology follows $Q$ or the shape of the kernel. To separate them we simulated kernels and parameters in which $Q$ and $q$ are not correlated: heavy-tailed kernels with fixed median radius (N2), fixed $\kappa$ (N3) and fixed core width (N4), the N1 sequence at $\beta=0.3$, and the Gaussian at other values of $D$, all at $\sigma=6\times10^{-3}$, together with a subset of these kernels at $\sigma=3\times10^{-3}$, which we list in the Supplemental Material together with the predicted stable states \cite{SM}. Figure~\ref{fig:collapse}(a) shows the skewness of the $49$ states against $Q$ evaluated at the simulated $\alpha$. The sign of the skewness follows the sign of $Q$ in every case in which $|Q|$ is not negligible, and the states with $Q\simeq0$ are labyrinths. At fixed $\sigma$, kernels as different as N2 and N3 at $q=1.8$ fall on the same curve, so the tail acts on the morphology only through $Q$. The curve, however, depends on $\sigma$. At $\sigma=3\times10^{-3}$ the gap branch saturates at a smaller $|Q|$. The N3 kernel with $q=1.8$, for instance, gives a skewness of $-0.21$ at $\sigma=6\times10^{-3}$ and of $-0.71$ at $\sigma=3\times10^{-3}$, with almost the same $Q$, while on the side of spots both values of $\sigma$ coincide.

This dependence on $\sigma$ is the one that the amplitude equations predict. Hexagons are stable for $\sigma<\sigma_H=Q^2(2g_1+g_2)/(g_2-g_1)^2$ and stripes for $\sigma>\sigma_S=Q^2g_1/(g_2-g_1)^2$, so the choice between hexagons and stripes depends on the ratio $\sigma/Q^2$, weighted by the cubic coefficients, and not on $Q$ alone. A smaller $\sigma$ widens the range of $Q$ in which only hexagons are stable. In Fig.~\ref{fig:collapse}(b) we group the $49$ states according to this prediction, which we evaluate with the coefficients at the simulated $\alpha$. The $16$ states for which only hexagons are stable have $|\mathrm{skewness}|\ge0.71$ with the sign of $Q$, i.e. they are spots or gaps. The $8$ states for which only stripes are stable have $|\mathrm{skewness}|\le0.29$ and are labyrinths. The $10$ states in the bistable range spread from labyrinths to strong spots or gaps, as we expect when both states are stable and the random initial condition selects one of them. Finally, the $15$ states with $g_1<0$, where the truncation does not apply, are gaps in every case. The amplitude equations then predict the morphology of every state in which the cubic truncation holds, and beyond it the sign of $Q$ still decides between spots and gaps.

\subsection{Biomass shift}\label{sec:biomass_sim}

\begin{figure}[t]
\includegraphics[width=\columnwidth]{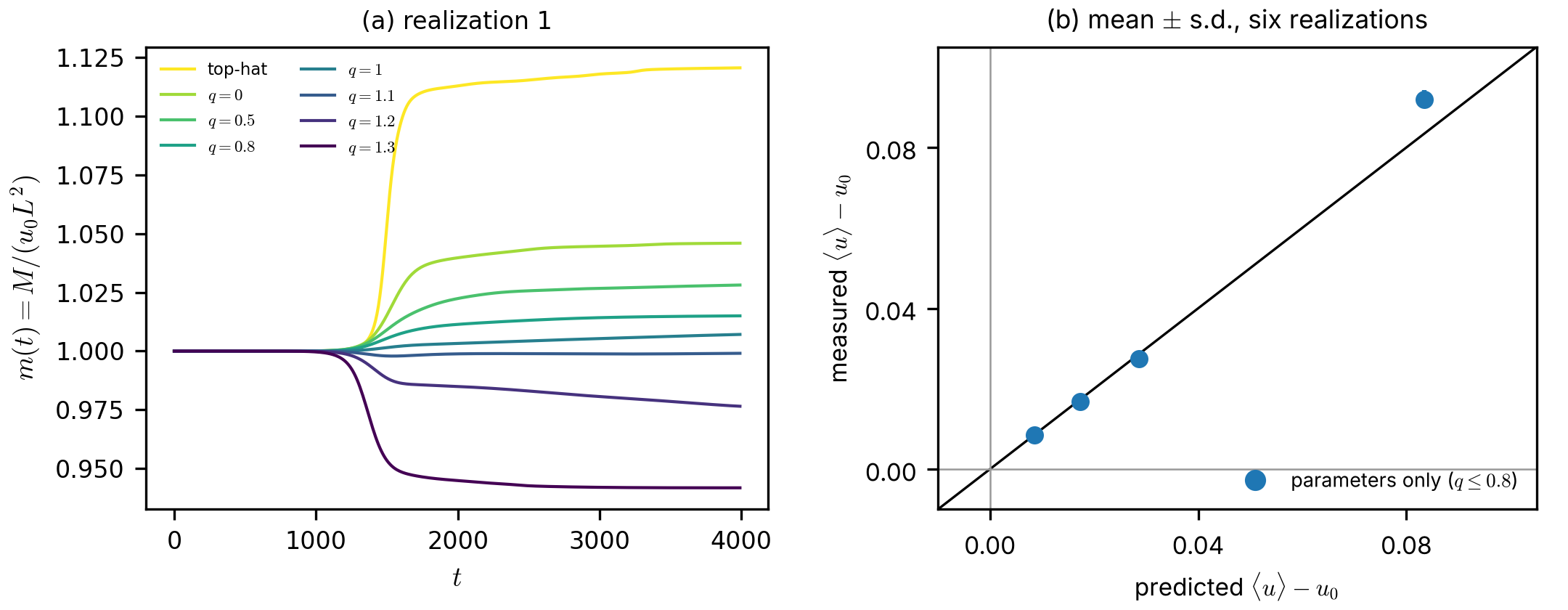}
\caption{(a) Total biomass $m(t)=M/(u_0L^2)$ for the different kernels (realization 1). (b) Measured shift $\avg u-u_0$ (mean $\pm$ s.d. over six realizations) against the prediction from the parameters alone, Eq.~\eqref{eq:biomass}, for $q\le0.8$. The diagonal is the reference.}
\label{fig:biomass}
\end{figure}

As we can see in Fig.~\ref{fig:biomass}(a), the total biomass of the patterned state grows above $u_0L^2$ for the kernels that give spots and decreases below it for those that give gaps, in agreement with the sign of $Q$. For the kernels where the cubic truncation holds ($q\le0.8$), the prediction of Eq.~\eqref{eq:biomass} from the parameters alone lies within $10\%$ of the measured shift [Fig.~\ref{fig:biomass}(b)], with ratios $0.91$, $1.05$, $1.03$ and $1.00$ from the top-hat to $q=0.8$. This agreement, however, is partly fortuitous. The cubic truncation predicts a variance $1.2$ to $1.6$ times smaller than the measured one, while Eq.~\eqref{eq:biomass} evaluated with the measured variance overestimates the shift by a factor $1.3$ to $1.7$, and the two errors compensate. For $q\ge1$ the prediction is not meaningful, since either $Q\simeq0$ and the shift lies below the dispersion between realizations, or $g_1<0$. We therefore regard Eq.~\eqref{eq:biomass} as the mechanism that links the morphology with the biomass, and as a quantitative estimate only for spots far from the fold.

%==========================================================
\section{Discussion and conclusions}\label{sec:disc}
%==========================================================

We have analyzed a simple model of vegetation with a strong Allee effect and nonlocal competition, and we have followed the consequences of the local activation from the linear instability to the selected pattern and the total biomass. That a positive local feedback relaxes the requirement of a negative kernel transform was already known \cite{clerc2005,clerc2010,escaff2015,vandervoort2026}. What the present work adds is the identification of a single coefficient that organizes the nonlinear consequences of this feedback.

First, the local activation allows a finite-wavelength instability for any kernel of the $q$-Gaussian family, including those with $\hat G>0$ and infinite variance. The threshold depends on which length we hold fixed when we compare kernels, and the closure of the instability window that appears at fixed rms radius comes from the shrinking core and not from the tail. The decrease of $k_c$ as the tail becomes heavier appears for every normalization.

Second, the quadratic coefficient of the hexagonal amplitude equations, which we obtained in closed form and verified to better than $1\%$, results from the balance between the curvature of the local growth, which favors spots, and the competitive inhibition at the pattern scale, which favors gaps. Its sign depends on whether $\hat G(k_c)$ exceeds the kernel-independent value $G^*(\alpha)$, and at onset it reduces to the comparison of $Dk_c^2/\cA$ with $(u_0+\alpha)/(u_0+2\alpha)$, Eq.~\eqref{eq:onset}, in which the kernel enters only through $k_c$.

Third, spots, labyrinths and gaps appear with pairwise competition only, in contrast with competition-only models, in which stripes and gaps require higher-order competition \cite{pintoramos2026}. The sign of $Q$ decides between spots and gaps for different normalizations, mortality rates, diffusion coefficients and distances to threshold, and the amplitude equations predict whether hexagons or stripes appear in every simulated state in which the cubic truncation holds. The strength of the gaps follows $Q$ and $\sigma$, and not the tail exponent.

Fourth, the total biomass increases for spots and decreases for gaps, with a leading-order shift proportional to $Q$ and to the variance of the pattern.

There is however a caveat. The parameters for which the sequence of morphologies appears lie close to the fold of the homogeneous state ($\alpha_c/\alpha_s\gtrsim0.92$ for $\beta=0.15$), where extinction is also stable, $|\lambda(0)|$ is small, and the cubic truncation fails for the gap patterns. In this regime $Q$ still orders the morphology, but the amplitude equations do not predict the saturated state, and the biomass relation gives the sign of the shift but not its magnitude. The labyrinth--gap boundary is a crossover that depends on the domain size and on the integration time, and our results near it should be read as a trend. Part of our statistics also rests on two realizations, in the N2 kernels with $q=1.2$, $1.5$ and $1.7$ and in the runs at $\sigma=3\times10^{-3}$. It is also worth recalling that the patterns appear for $\alpha=c/K\simeq1.1$--$1.2$, which is admissible because $\alpha$ is not the Allee threshold, but which corresponds to a strongly cooperative growth. Our model establishes a mechanism in a minimal setting, and not evidence that this regime is typical of dryland vegetation. In the same way, biomass is a state variable of the model and we cannot identify it with productivity, since the model has no explicit resource dynamics.

The present results suggest some questions for future work. The amplitude equations predict the morphology wherever the cubic truncation holds, and the gaps persist, with a skewness close to that of an ideal hexagonal pattern, where it fails. A description of these saturated gaps beyond the cubic order is still missing. It would also be interesting to estimate, from root and water-redistribution data, the transform of real competition kernels at the pattern scale, since our results indicate that this value, and not the tail, is what matters for the morphology. Finally, models with explicit water dynamics or nonlocal facilitation would tell whether the balance between the local curvature and the competitive inhibition at the pattern scale controls the morphology there as well.

%==========================================================
\appendix
\section{Linearization and instability criterion}\label{app:lin}
%==========================================================

We write $F(u,\bar u)=h(u)(1-\bar u)-\beta u$ with $h(u)=u^2/(u+\alpha)$, so that
\begin{equation}
h'=\frac{u(u+2\alpha)}{(u+\alpha)^2},\quad
h''=\frac{2\alpha^2}{(u+\alpha)^3},\quad
h'''=-\frac{6\alpha^2}{(u+\alpha)^4}.
\end{equation}
The partial derivatives are
\begin{align}
F_u&=h'(1-\bar u)-\beta,& F_{\bar u}&=-h,\nonumber\\
F_{uu}&=h''(1-\bar u),& F_{u\bar u}&=-h',\nonumber\\
F_{uuu}&=h'''(1-\bar u),& F_{uu\bar u}&=-h'',
\end{align}
and every derivative with two or more $\bar u$ vanishes, because $F$ is linear in $\bar u$. At a homogeneous state we use $\beta=u_0(1-u_0)/(u_0+\alpha)$ to eliminate $\beta$, and we obtain
\begin{widetext} 
\begin{align}
\cA&\equiv F_u(u_0,u_0)=\frac{u_0(1-u_0)}{(u_0+\alpha)^2}\big[(u_0+2\alpha)-(u_0+\alpha)\big]=\frac{\alpha\beta}{u_0+\alpha},\nonumber\\
\cB&\equiv F_{\bar u}(u_0,u_0)=-\frac{u_0^2}{u_0+\alpha}.
\end{align}
\end{widetext}
A perturbation $u=u_0+\varepsilon\phi$ obeys, to first order, $\partial_t\phi=D\nabla^2\phi+\cA\phi+\cB\,G*\phi$. For $\phi=\hat\phi e^{i\kk\cdot\rr}$ the Laplacian gives $-k^2$ and the convolution theorem gives $G*\phi=\hat G(k)\phi$, which yields Eq.~\eqref{eq:lambda}. Multiplying $\lambda(k)>0$ by $u_0+\alpha>0$ we obtain Eq.~\eqref{eq:criterion}.

For uniform perturbations we define the per-capita rate $f(u)=u(1-u)/(u+\alpha)-\beta=-(u-u_0^+)(u-u_0^-)/(u+\alpha)$, so that the local model is $\dot u=uf(u)$ and $\lambda(0)=u_0f'(u_0)$, because $f(u_0)=0$. Since $f'(u_0^\pm)=\mp(u_0^+-u_0^-)/(u_0^\pm+\alpha)$, for $u_0^+$ we obtain Eq.~\eqref{eq:lambda0}, while for $u_0^-$ the sign is positive.

Three consequences follow from $\lambda=-Dk^2+\cA+\cB\hat G$. (i)~If $\cA=0$ (no Allee effect, $\alpha\to0$), $\lambda>0$ requires $\hat G<0$. (ii)~If $\hat G\ge0$, $\lambda>0$ requires $\cA>Dk^2+|\cB|\hat G(k)$, and then $k<\sqrt{\cA/D}$. Long wavelengths are stable because $|\cB|\hat G(0)=|\cB|$ exceeds $\cA$, and short ones because of diffusion. In between, $\hat G(k)$ decays and can fall below $\cA/|\cB|$ while $Dk^2$ is still small, which opens the unstable band whatever the sign of $\hat G$. (iii)~If $\hat G(k)\simeq1$ for $k\lesssim\sqrt{\cA/D}$ (a kernel narrower than $\sqrt{D/\cA}$), then $\lambda(k)\simeq\lambda(0)-Dk^2<0$ for all $k$, so a nearly local kernel does not destabilize the homogeneous state. This is the origin of the closure of the window with N1 as $q\to3/2$.

For the competition-only model $\partial_t\rho=\beta_0\rho(1-\rho)/(1+\delta\bar\rho)-\gamma\rho$ of Ref.~\cite{martinezgarcia2013}, the homogeneous state satisfies $\beta_0(1-\rho_0)/(1+\delta\rho_0)=\gamma$, and we find $F_{\bar\rho}=-\delta\gamma\rho_0/(1+\delta\rho_0)$ and $F_\rho=-\gamma\rho_0/(1-\rho_0)<0$. The local term is an inhibition, the dispersion relation without diffusion is $\lambda(k)=\gamma\rho_0[-\delta\hat G/(1+\delta\rho_0)-1/(1-\rho_0)]$, and $\lambda>0$ requires $\hat G<0$. With $\rho_0=(\beta_0-\gamma)/(\beta_0+\delta\gamma)$ the condition $\lambda>0$ becomes $(\delta\gamma|\hat G|_{\min}-\beta_0)(1+\delta)>0$, i.e. $\delta_c=\beta_0/(\gamma|\hat G|_{\min})$.

%==========================================================
\section{Fourier transforms and normalizations}\label{app:ft}
%==========================================================

For a radial function the $d$-dimensional Fourier transform reduces to a Hankel transform,
\begin{equation}
\hat G(k)=(2\pi)^{d/2}k^{1-d/2}\int_0^\infty G(r)\,J_{d/2-1}(kr)\,r^{d/2}\,dr .
\end{equation}

\paragraph{Compact kernels, $q<1$.} We write $n=1/(1-q)$ and $G(r)=C^{-1}(1-r^2/R^2)_+^n$, with $C=R^d\pi^{d/2}\Gamma(n+1)/\Gamma(n+d/2+1)$ so that $\int G=1$. Sonine's first finite integral, $\int_0^1J_\mu(kr)\,r^{\mu+1}(1-r^2)^\nu dr=2^\nu\Gamma(\nu+1)\,k^{-\nu-1}J_{\mu+\nu+1}(k)$, with $\mu=d/2-1$ and $\nu=n$ \cite{watson1944,gradshteyn2007}, gives
\begin{equation}
\int_{|x|<1}\!(1-|x|^2)^ne^{-i\kk\cdot x}d^dx=\pi^{d/2}\Gamma(n+1)\Big(\frac2k\Big)^{n+\frac d2}\!J_{n+\frac d2}(k).
\end{equation}
Dividing by the value at $k=0$ and rescaling $k\to kR$ we obtain Eq.~\eqref{eq:Gq_compact} with $\nu=n+d/2$. For $d=2$ and $n=0$ it reduces to $2J_1(k)/k$ (top-hat).

\paragraph{Power-law kernels, $q>1$.} We write $s=1/(q-1)$ and $G=C^{-1}(1+r^2/a^2)^{-s}$, with $C=a^d\pi^{d/2}\Gamma(s-d/2)/\Gamma(s)$. Basset's integral, $\int_0^\infty x^{\nu+1}J_\nu(bx)(x^2+a^2)^{-(\rho+1)}dx=a^{\nu-\rho}b^{\rho}K_{\nu-\rho}(ab)/[2^\rho\Gamma(\rho+1)]$, with $\nu=d/2-1$ and $\rho=s-1$, shows that the transform is a Mat\'ern function. After normalization to $\hat G(0)=1$, using $x^\mu K_\mu(x)\to2^{\mu-1}\Gamma(\mu)$ as $x\to0$, it is Eq.~\eqref{eq:Gq_tail} with $\mu=s-d/2$. The kernel is normalizable if $s>d/2$, i.e. $q<1+2/d$. For $d=1$ and $q=2$, $K_{1/2}(z)=\sqrt{\pi/2z}\,e^{-z}$ gives $\hat G=e^{-ka}$, the transform of the Lorentzian. We checked both formulas against direct numerical integration of the Hankel transform (agreement to five digits).

\paragraph{Normalizations.} In $d=2$ the normalized kernels are $G=\frac{n+1}{\pi R^2}(1-r^2/R^2)_+^n$ and $G=\frac{s-1}{\pi a^2}(1+r^2/a^2)^{-s}$. With $t=r^2/R^2$ (or $r^2/a^2$) the radial integrals become Beta functions. The second moments are $\avg{r^2}=R^2/(n+2)$ and $\avg{r^2}=a^2/(s-2)$, finite only for $s>2$, i.e. $q<3/2$, and fixing $\avg{r^2}=\ell^2$ (N1) gives the expressions of Sec.~\ref{sec:qgauss}. In $d=2$, $a=\ell\sqrt{s-2}\to0$ as $q\to3/2$, so the variance constraint forces the core of the kernel to shrink. The mass enclosed in a radius $r$ is $1-(1-r^2/R^2)^{n+1}$ and $1-(1+r^2/a^2)^{1-s}$, respectively, and setting it to $1/2$ gives the N2 expressions, finite for every normalizable kernel ($q<2$). For $q\to1$ all the families tend to $G=(\kappa/\pi)e^{-\kappa r^2}$, with $\hat G=e^{-k^2/4\kappa}$, $\avg{r^2}=1/\kappa$ and $r_{1/2}^2=\ln2/\kappa$, so that with $\kappa=1/\ell^2$ and $r_{1/2}=\sqrt{\ln2}\,\ell$ the three normalizations coincide at $q=1$. The fixed-core family N4 uses $a=1$.

%==========================================================
\section{Derivation of the amplitude equations}\label{app:wnl}
%==========================================================

\paragraph{Expansion.} With $\phi=u-u_0$ and $\psi=G*\phi$, the Taylor expansion of $F(u_0+\phi,u_0+\psi)$ about $(u_0,u_0)$ gives $\partial_t\phi=\mathcal L\phi+N_2+N_3$, where $\mathcal L\phi=D\nabla^2\phi+\cA\phi+\cB\,G*\phi$ acts on a Fourier mode as a multiplication by $\lambda(k)$, and $N_2$, $N_3$ are Eq.~\eqref{eq:N23}. The derivatives of Appendix~\ref{app:lin} give Eq.~\eqref{eq:ab}: $a_2=\tfrac12h''(u_0)(1-u_0)$, $b_2=-h'(u_0)$, $a_3=\tfrac16h'''(u_0)(1-u_0)$ and $b_3=-\tfrac12h''(u_0)$.

\paragraph{Quadratic terms on Fourier modes.} With $\phi=\sum_{\mathbf p}c_{\mathbf p}e^{i\mathbf p\cdot\rr}$ and $c_{-\mathbf p}=\bar c_{\mathbf p}$, we have $\phi\,(G*\phi)=\sum_{\mathbf p,\mathbf q}c_{\mathbf p}c_{\mathbf q}\hat G(q)e^{i(\mathbf p+\mathbf q)\cdot\rr}$. Grouping the two orderings of each pair of different wavevectors, the amplitude of $N_2$ at $\kk$ is
\begin{equation}
[N_2]_{\kk}=\sum_{\{\mathbf p,\mathbf q\}}\omega_{\mathbf p\mathbf q}\,W(\mathbf p,\mathbf q)\,c_{\mathbf p}c_{\mathbf q},
\label{eq:Wdef}
\end{equation}
over unordered pairs with $\mathbf p+\mathbf q=\kk$, with $W(\mathbf p,\mathbf q)=a_2+\tfrac12b_2[\hat G(p)+\hat G(q)]$, $\omega_{\mathbf p\mathbf q}=2$ for $\mathbf p\ne\mathbf q$ and $\omega=1$ otherwise. For two critical modes $W_c=a_2+b_2\hat G(k_c)=Q/2$, and $W_j=W(k_c,jk_c)$ couples a critical mode with a mode of wavenumber $jk_c$.

\paragraph{Slaved modes.} With $A_j=O(\varepsilon)$ and $\sigma=O(\varepsilon^2)$, the non-critical modes are $O(\varepsilon^2)$ and obey $\dot c_{\kk}=\lambda(k)c_{\kk}+[N_2]_{\kk}$. Their time derivative is of higher order, so $c_{\kk}=-[N_2]_{\kk}/\lambda(k)$ (adiabatic elimination). At $\kk=0$ the pairs $(\kk_j,-\kk_j)$ give $c_0=-\frac{2W_c}{\lambda_0}\sum_j|A_j|^2$. At $\kk=2\kk_j$ the pair $(\kk_j,\kk_j)$ gives $c_{2\kk_j}=-\frac{W_c}{\lambda_2}A_j^2$. At $\kk=\kk_i-\kk_j$ the pair $(\kk_i,-\kk_j)$ gives $c_{\kk_i-\kk_j}=-\frac{2W_c}{\lambda_{\sqrt3}}A_i\bar A_j$, since $|\kk_i-\kk_j|^2=3k_c^2$ for vectors at $120^\circ$.

\paragraph{Projection on the critical mode.} The terms proportional to $e^{i\kk_1\cdot\rr}$ are the linear term $\sigma A_1$; the resonant quadratic term, from the pair $(-\kk_2,-\kk_3)$, $2W_c\bar A_2\bar A_3=Q\bar A_2\bar A_3$; the direct cubic terms, from the ordered triples $(\kk_1,\kk_j,-\kk_j)$, which give $P[3|A_1|^2+6(|A_2|^2+|A_3|^2)]A_1$ with $P=a_3+b_3\hat G(k_c)$, where $3$ and $6$ count the orderings and the wavevector inside the convolution always has modulus $k_c$; and the quadratic terms with one critical and one slaved mode, $2W_0c_0A_1$, $2W_2c_{2\kk_1}\bar A_1$ and $2W_{\sqrt3}\sum_{j=2,3}c_{\kk_1-\kk_j}A_j$. Substituting the slaved amplitudes and collecting the coefficients of $|A_1|^2A_1$ and $(|A_2|^2+|A_3|^2)A_1$ we obtain Eq.~\eqref{eq:g}.

\paragraph{Swift--Hohenberg check.} For $\partial_tu=ru-(1+\nabla^2)^2u+gu^2-u^3$ we have $a_2=g$, $a_3=-1$, $b_2=b_3=0$, $\lambda_0=-1$, $\lambda_{\sqrt3}=-4$ and $\lambda_2=-9$, and Eq.~\eqref{eq:g} gives $g_1=3-38g^2/9$, the known value, and $g_2=6-5g^2$.

\paragraph{Ordering.} With $A=O(\varepsilon)$ and $\sigma=O(\varepsilon^2)$, the term $Q\bar A_2\bar A_3$ is $O(\varepsilon^2)$, so the expansion is formally consistent only for $Q=O(\varepsilon)$. We retain $Q$ at its actual value, and the quantitative check of Table~\ref{tab:verif} uses the branch $A\simeq-\sigma/Q$, which is small for small $\sigma$ whatever $Q$.

%==========================================================
\section{Solutions of the amplitude equations}\label{app:amp}
%==========================================================

\paragraph{Phase.} Writing $A_j=\mathcal R_je^{i\theta_j}$ and $\Theta=\theta_1+\theta_2+\theta_3$, Eq.~\eqref{eq:amp} gives
\begin{align}
\dot{\mathcal R}_1&=\sigma\mathcal R_1+Q\mathcal R_2\mathcal R_3\cos\Theta-g_1\mathcal R_1^3-g_2(\mathcal R_2^2+\mathcal R_3^2)\mathcal R_1,\nonumber\\
\dot\Theta&=-Q\sin\Theta\Big(\frac{\mathcal R_2\mathcal R_3}{\mathcal R_1}+\frac{\mathcal R_1\mathcal R_3}{\mathcal R_2}+\frac{\mathcal R_1\mathcal R_2}{\mathcal R_3}\Big),
\end{align}
so $\Theta=0$ is stable for $Q>0$ (spots) and $\Theta=\pi$ for $Q<0$ (gaps).

\paragraph{Hexagons.} For equal amplitudes the signed amplitude $A$ obeys $\sigma A+QA^2-gA^3=0$, with $g=g_1+2g_2$ and roots $A_\pm=[Q\pm\sqrt{Q^2+4g\sigma}]/(2g)$, real for $\sigma\ge-Q^2/4g$. Since $\frac{d}{dA}(\sigma A+QA^2-gA^3)|_{A_\pm}=\mp A_\pm\sqrt{Q^2+4g\sigma}$, for $g>0$ the branch $A_+$, which has the sign of $Q$, is stable. For $\sigma<0$ the branch $A_-$ has the sign of $Q$ and a smaller amplitude, and it is unstable in amplitude. For $\sigma>0$ it has the opposite sign, which corresponds to the unstable value of $\Theta$. Near threshold $A_-\simeq-\sigma/Q$ and $A_+\simeq Q/g$. Hexagons are stable against stripes for $\sigma<Q^2(2g_1+g_2)/(g_2-g_1)^2$ \cite{cross1993,cross2009}.

\paragraph{Stripes.} With $A_1=\mathcal R$ and $A_2=A_3=0$ the quadratic term vanishes and $\mathcal R^2=\sigma/g_1$, which requires $g_1>0$. Linearizing for small $A_2$, $\bar A_3$ gives $\dot A_2=(\sigma-g_2\mathcal R^2)A_2+Q\mathcal R\bar A_3$ and $\dot{\bar A}_3=(\sigma-g_2\mathcal R^2)\bar A_3+Q\mathcal RA_2$, with eigenvalues $\sigma(1-g_2/g_1)\pm|Q|\sqrt{\sigma/g_1}$. Both are negative if and only if $g_2>g_1$ and $\sigma>Q^2g_1/(g_2-g_1)^2$.

\paragraph{Moments.} For a hexagonal pattern $\phi=2A\sum_j\cos(\kk_j\cdot\rr)$, $\avg{\phi^2}=6A^2$ and $\avg{\phi^3}=12A^3$, so the skewness is $\mathrm{sign}(A)\,2/\sqrt6\simeq\pm0.816$, which we confirmed numerically. A stripe pattern has $\avg{\phi^3}=0$. The measured skewness of the weak spot patterns ($0.78$ at $q=0.8$) is of this order, and the larger values of the top-hat reflect the strongly nonlinear regime.

%==========================================================
\section{Biomass relation}\label{app:bio}
%==========================================================

Integrating Eq.~\eqref{eq:model} over the periodic domain $\Omega$ of area $L^2$, the diffusion term vanishes and $dM/dt=\int_\Omega F\,d\rr$, so $\avg F=0$ in a stationary state, exactly and for any amplitude. Let $\bar m=\avg u$ and $\delta u=u-\bar m$, with $\avg{G*\delta u}=\avg{\delta u}=0$ since $\int G=1$. Expanding $F$ about $(\bar m,\bar m)$ and averaging, the linear terms vanish and $F(\bar m,\bar m)+a_2\avg{\delta u^2}+b_2\avg{\delta u\,(G*\delta u)}+\dots=0$. We linearize $F(\bar m,\bar m)\simeq\lambda(0)(\bar m-u_0)$, because $\frac{d}{dm}F(m,m)=F_u+F_{\bar u}=\lambda(0)$ at $u_0$, evaluate $a_2$ and $b_2$ at $u_0$, and keep a single ring, so that $\avg{\delta u\,(G*\delta u)}=\hat G(k_c)\avg{\delta u^2}$. We obtain $\lambda(0)(\bar m-u_0)+[a_2+b_2\hat G(k_c)]\avg{\delta u^2}=0$, which is Eq.~\eqref{eq:biomass}. The same result follows from the slaved zero mode of Appendix~\ref{app:wnl}, $c_0=Q\sum_j|A_j|^2/|\lambda_0|$, with $\avg{\delta u^2}=2\sum_j|A_j|^2$. The linearization requires $|\bar m-u_0|\ll u_0^+-u_0^-$, since $F(m,m)=-m(m-u_0^+)(m-u_0^-)/(m+\alpha)$ varies on that scale, and it fails near the fold.

%==========================================================
\section{Numerical methods and observables}\label{app:num}
%==========================================================

\paragraph{Time integration.} Treating the diffusion term implicitly, $(\hat u^{n+1}-\hat u^n)/\Delta t=-Dk^2\hat u^{n+1}+\hat F^n$, we obtain Eq.~\eqref{eq:scheme}. The scheme is first order in $\Delta t$ and unconditionally stable for the diffusion, and the explicit reaction term limits $\Delta t$ by a rate of order one, so $\Delta t=0.25$ lies well within the stable range. We obtain $\bar u$ by multiplying $\hat u$ by $\hat G(|\kk|)$ at $\kk=2\pi\mathbf n/L$, which by the Poisson summation formula is equivalent to convolving with the periodized kernel $\sum_{\mathbf j}G(\rr+L\mathbf j)$. We clip $u\ge0$ as a safeguard, and in the runs that we monitored no pixel became negative.

\paragraph{Choice of $\alpha$.} For each kernel we solve $\max_k\lambda(k;\alpha)=6\times10^{-3}$ for $\alpha$ with Brent's method, and $k_c$ is the position of the maximum at that $\alpha$. We obtain the thresholds $\alpha_c$ in the same way with $\max_k\lambda=0$.

\paragraph{Euler characteristic.} We represent the set $\{u>\avg u\}$ as a union of closed pixels on the periodic grid and compute $\chi=V-E+F$, with $F$ the number of pixels in the set and $E$ and $V$ the edges and vertices that belong to at least one of them \cite{adler2007}. For $n_s$ isolated components $\chi=n_s$, and for a connected set with $n_h$ holes $\chi=1-n_h$. The variant with open pixels differs by at most $0.002$ in $\chi/L^2$.

\paragraph{Structure factor and hexatic order.} We compute $S=L^{-2}|\hat u|^2$ from the FFT of $\delta u$, average it over annuli of width $\Delta k=2\pi/L$, and take $k^*$ as the $S$-weighted mean of the five bins around the maximum. We detect maxima (or minima) as the points equal to the maximum of their $5\times5$ periodic neighbourhood and above (below) the mean, merging plateaus. A Delaunay triangulation of the $3\times3$ periodic images gives the neighbours for $\psi_6(j)=n_j^{-1}\sum_le^{6i\theta_{jl}}$, and for uniformly random points $\overline{|\psi_6|}\simeq0.36$.

%==========================================================
\section{Fourier--Galerkin verification}\label{app:verif}
%==========================================================

We computed the stationary states with the symmetry of a stripe or of a hexagonal lattice of wavenumber $k_c$ with a Fourier--Galerkin method. For stripes $u(x)=c_0+2\sum_{n=1}^{6}c_n\cos(nk_cx)$, and for hexagons $u=\sum_{\mathcal O}c_{\mathcal O}\sum_{\kk\in\mathcal O}e^{i\kk\cdot\rr}$, where $\mathcal O$ are the orbits of the lattice $\{m\kk_1+n\kk_2\}$ under the dihedral group $D_6$ with $m^2-mn+n^2\le12$. We evaluate the nonlinear term pseudospectrally on a $64\times36$ grid of the cell $L_x=4\pi/k_c$, $L_y=4\pi/(\sqrt3k_c)$, and solve the residual projected on the orbit representatives with Newton's method. We vary $\alpha$ above $\alpha_c$ with the lattice wavenumber fixed at $k_c(\alpha_c)$, and evaluate the coefficients at the same $(\alpha,k_c)$. For stripes we extrapolate $\sigma/A^2$ linearly in $A^2$ to $\sigma\to0$, which gives $g_1$. For hexagons we use the small-amplitude branch, $A\simeq-\sigma/Q$, and extrapolate $(\sigma+QA)/A^2$ linearly in $A$, which gives $g_1+2g_2$. For the Gaussian ($Q\simeq0$) we use instead the stable branch, after relaxing the field by time stepping on the same cell.

%==========================================================

\end{document}